%% file: main.tex
\documentclass[runningheads]{llncs}
\usepackage[pdftex]{graphicx}
\usepackage[T1]{fontenc}
\usepackage{graphicx}
\usepackage[english]{babel}\addto\captionsenglish  
{}  
\usepackage{blindtext}
\usepackage{tabularx}
\usepackage[utf8]{inputenc}
\usepackage[caption=false]{subfig}
\usepackage{graphicx}
\usepackage{mathtools}
\usepackage{pifont}
\usepackage{comment}

\usepackage{mathtools}
\usepackage{authblk}

\usepackage{amsthm}
\usepackage{amsmath}
\usepackage{xcolor}
\usepackage[linesnumbered,ruled,vlined]{algorithm2e}

\usepackage{tabulary} 
\usepackage{booktabs} 
\usepackage{multirow} 
\usepackage{array} 
\usepackage{ragged2e}

\usepackage{amsmath}       
\usepackage{nicematrix}    

\usepackage{amsthm}
\usepackage{float}
\usepackage{csquotes}
\usepackage[hidelinks]{hyperref}

\begin{document}
\pagenumbering{arabic}
\setcounter{page}{1}
\input{Sections/0_TitleAbstract}

    \input{Sections/1_Introduction}

     \input{Sections/2_ProblemDescription}

    \input{Sections/3_ProblemModel}

    \input{Sections/4_ProposedMethod}
    \input{Sections/5_Evaluation}

    \input{Sections/6-related2}
    \input{Sections/7_Conclusion}

\bibliography{myref}

\end{document}

%% file: Sections/0_TitleAbstract.tex
\title{Reinforcement Learning-Driven Adaptation Chains: A Robust Framework for Multi-Cloud Workflow Security}

\titlerunning{A Robust Framework for Multi-Cloud Workflow Security}

\author{Nafiseh Soveizi\inst{1}\orcidID{0000-0003-2111-734X} \and
Dimka Karastoyanova\inst{1}\orcidID{0000-0002-8827-2590}}
\authorrunning{N.Soveizi and D.Karastoyanova}
%
\institute{Information Systems Group, University of Groningen, Groningen, The Netherlands
\email{\{n.soveizi,d.karastoyanova\}@rug.nl}}
\maketitle            


\begin{abstract}
Cloud computing has emerged as a crucial solution for managing data- and compute-intensive workflows, offering scalability to address dynamic demands. However, security concerns persist, especially for workflows involving sensitive data and tasks. One of the main gaps in the literature is the lack of robust and flexible measures for reacting to these security violations. To address this, we propose an innovative approach leveraging Reinforcement Learning (RL) to formulate adaptation chains, responding effectively to security violations within cloud-based workflows. These chains consist of sequences of adaptation actions tailored to attack characteristics, workflow dependencies, and user-defined requirements. Unlike conventional single-task adaptations, adaptation chains provide a comprehensive mitigation strategy by taking into account both control and data dependencies between tasks, thereby accommodating conflicting objectives effectively. Moreover, our RL-based approach uses insights from past responses to mitigate uncertainties associated with adaptation costs. We evaluate the method using our jBPM and Cloudsim Plus based implementation and compare the impact of selected adaptation chains on workflows with the single adaptation approach. Results demonstrate that the adaptation chain approach outperforms in terms of total adaptation cost, offering resilience and adaptability against security threats.

\keywords{Security-aware workflows  \and Cloud-based workflows \and Workflow Adaptation \and Cloud Service Monitoring \and Adaptation Recommendation \and Adaptation Chain}
\end{abstract}

%% file: Sections/1_Introduction.tex
\section{Introduction}
\vspace{-10pt}
Cloud computing has become an essential solution for organizations managing data- and compute-intensive workflows, offering unparalleled scalability and flexibility to meet dynamic demands \cite{sooezi2015scheduling}. This transformative shift enables organizations to outsource workflow execution, fundamentally altering operational approaches. However, despite the numerous benefits of cloud-based workflows, concerns about cloud security remain prominent \cite{Varshney2019,Maguluri2012}, particularly impeding the adoption of such workflows involving sensitive data and tasks.

The dynamic binding of workflows to cloud services introduces significant security concerns due to increased security risks and malicious attacks \cite{soveizi2023security}. Security concerns are further compounded by the transmission of sensitive data among cloud components, such as Data Centers (DCs), over potentially untrusted network channels. Hence, there exists a critical necessity to establish a systematic approach capable of responding to such attacks and threats in a flexible way.

While there are efforts that address the challenge of detecting attacks and security breaches within cloud-based workflows and responding to identified threats with workflow adaptation \cite{soveizi2023enhancing,Wang2020bb,poola2017taxonomy,Alaei2021}, we acknowledge the limitations inherent to such approaches. These limitations include: 1) \textit{Insufficient Mitigation of Attack Impact}: Single-task adaptation may not fully address the overall impact of an attack, neglecting control and data dependencies between tasks. Considering potential adaptations in other workflow tasks through an understanding of control and data dependencies could provide more comprehensive coverage of the attack's impact. 2) \textit{Optimized Trade-offs and Tailored Customization}: In scenarios with multiple objectives, a Single-task adaptation strategy cannot concurrently address conflicting goals. 3) \textit{Effective Handling of Complex Attack Scenarios}: In complex attack scenarios, where detection may experience delays (occurring when the engine processes the remaining tasks in the workflow), relying on a single-task adaptation strategy poses significant risks and costs.

In response to these challenges, this paper proposes a novel approach that aims at identifying chains of adaptation actions that would address the above-mentioned challenges.  We employ Reinforcement Learning (RL) to derive a sequence of adaptation actions, termed an "adaptation chain", which is similar to the concept of adaptation processes that defines steps of workflow activities to carry out (typically control-flow, runtime) adaptations \cite{reichert2012enabling,reichert2009flexibility}. In our work, we aim to tailor the adaptation chains to the attack's characteristics, the workflow's nature, and user-defined requirements. Unlike single-task adaptation constraints, adaptation chains ensure a more holistic mitigation approach by considering control and data dependencies between tasks. They offer a balanced response to various considerations in reacting to security violations, including violation severity and user preferences. Additionally, adaptation chains prove effective in managing multi-pronged attacks, addressing the complexity of sophisticated scenarios efficiently. This strategy enhances the resilience and robustness of the system by integrating multiple adaptations and increases the overall flexibility of processes.

We introduce the approach along the following lines: In Section 2, we present a motivating example to illustrate the problem. Section 3 provides an overview of the workflow and system model, along with foundational definitions of the problem. In Section 4, we detail the proposed approach to identify the optimal adaptation chain. Section 5 presents our evaluation setup and results. Section 6 provides an overview of the related work. Section 7 concludes the paper and outlines potential future research directions.

%% file: Sections/2_ProblemDescription.tex
\section{Motivating scenario}
\label{sec:Problem Description}
\begin{figure*}[htbp]
\vspace{-14pt}
  \includegraphics[width=\textwidth]{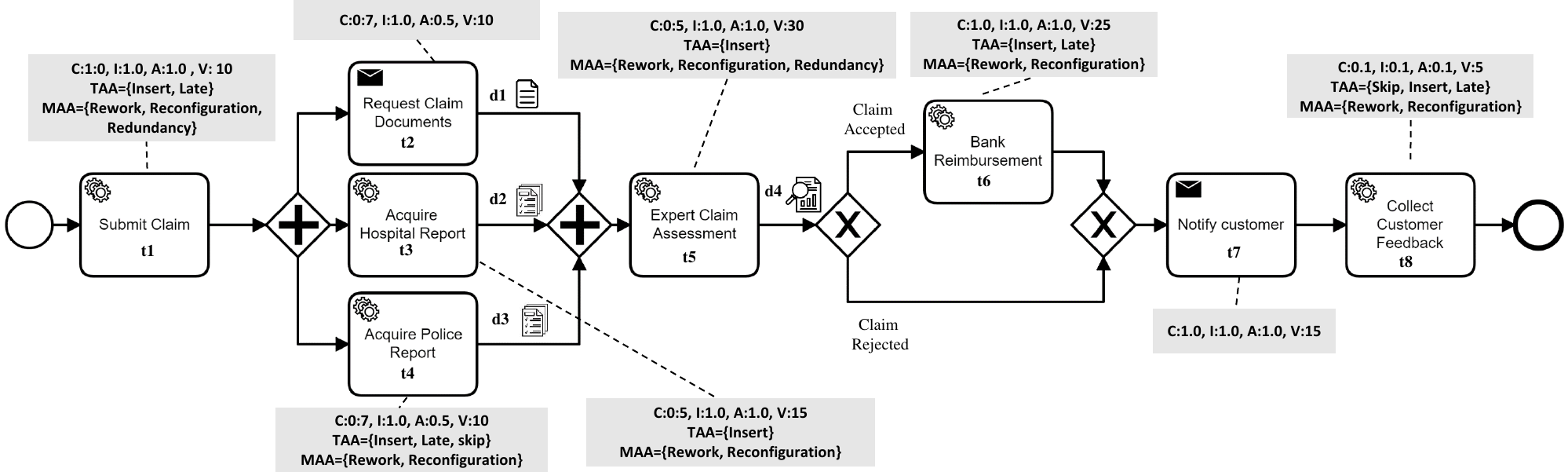}
  \caption{BPMN Workflow of Insurance Claim Recovery Process}
  \captionsetup{skip=0pt}
  \label{fig:WorkflowExample}
  \vspace{-14pt}
\end{figure*}

\vspace{-10pt}
\label{subsec:Motivation scenario}
In Figure \ref{fig:WorkflowExample} we present a BPMN workflow depicting a common insurance claim recovery process \cite{Goettelmann2015a}, which serves as our running example and supports our motivation. The workflow comprises eight tasks (t1-t8), each integral to the process. The process starts with an insurance company receiving a claim recovery declaration ("Submit Claim") from a customer, which includes sensitive customer information such as name, address, and bank account details. Subsequently, tasks including "Acquire Hospital Report", "Acquire Police Report", and "Request Claim Documents" are initiated, acting as prerequisites for the "Expert Claim Assessment" task. In this step, designated experts assess the claim according to a predefined protocol. Based on the experts' assessment, the "Bank Reimbursement" process may be activated. Regardless of the assessment outcome (negative or positive), the customer is informed through the "Notify customer" task. Finally, the "Collect Customer Feedback" task compiles customer feedback, facilitating the assessment of their experiences throughout the claim recovery process.

Table \ref{table:Attack_Scenarios} shows examples of possible attack scenarios that might occur while executing the workflow from Figure \ref{fig:WorkflowExample}.

\begin{table*}[ht]
\vspace{-14pt}
\scriptsize
\fontsize{7pt}{8pt}\selectfont
\caption{Attack Scenarios in BPMN Scenario}
\label{table:Attack_Scenarios}
\def\arraystretch{1}
\centering 
 \input{tables/table1-v3}
 \vspace{-14pt}
\end{table*}


%% file: tables/table1-v3.tex
\begin{tabulary}{\linewidth}{>{\arraybackslash}p{1.8cm}|>{\arraybackslash}p{2.2cm}|>{\arraybackslash}p{2.4cm}|>{\arraybackslash}p{5.5cm}}
 \hline
\centering\textbf{Attack Type} & \centering\textbf{Targeted Assets} & \centering\textbf{Malicious Party} &\textbf{Attack Scenario} \\
\hline \hline
DOS \linebreak (Denial \linebreak of Service)  & Task (t1) & Malicious Tenant/ External Attacker & A coordinated Denial of Service (DOS) attack orchestrated by a rival company to disrupt the claims processing system. \\
\hline
R2L \linebreak (Remote-to-Local) & Task (t3, t4)/ Intermediate data (d2, d3, d4) & Malicious Workflow User/ Malicious Tenant&  Unauthorized access to manipulate incident details or expert assessments for malicious purposes.\\
\hline
U2R \linebreak (User-to-Root) & Task (t5)/ Intermediate data (d4) & Malicious Workflow User& Exploits vulnerabilities to gain unauthorized access to alter reimbursement decisions. \\
\hline
Probe  & Tenant’s Metadata/ Users’ Metadata/ Logic & Malicious Provider (Semi-Trusted)/ Malicious Tenant& Gain access to observe process logic, decision criteria, and involved parties, accumulating intelligence about the reimbursement process, potentially for misuse or, more broadly, to access sensitive data within the process, such as customer information.\\
\hline 
\bottomrule 
\end{tabulary}\par 

%% file: Sections/3_ProblemModel.tex
\section{Problem Model}
\vspace{-10pt}
\label{sec:Problem Model}
In this section we will present a novel workflow adaptation strategy to react to security violations in cloud-based environments that extends the previous method by \cite{soveizi2023enhancing} and \textit{SecFlow} (\cite{soveizi2023secflow}), which are choosing a single adaptation action in response to detected attacks. This new approach involves a sequence of adaptation actions, referred to as an \textit{"adaptation chain"} that defines the steps to be taken to adapt a running workflow in response to a security violation. 
The adaptation chain is constructed dynamically upon a security violation by solving an Optimal Adaptation Chain Selection (OACS) problem. This problem is defined in the remainder of this section.
\vspace{-8pt}
\subsection{Workflow and System Models}
\vspace{-8pt}
In this section, we outline the foundational elements of the workflow and system models, as introduced by \cite{soveizi2023enhancing}, essential for defining the OACS problem. 

\label{subsection:Definitions}

\vspace{-5pt}
\begin{definition}\label{def:1}
(\textit{Workflow}).
\normalfont A workflow $w$ consists of abstract Service Tasks ($ST$), intermediate Data ($D$) exchanged between tasks, control Edges ($E_c$) determining task execution order based on specified conditions, and data Edges ($E_d$) specifying data flow between tasks.
\end{definition}

\vspace{-5pt}

\begin{definition}\label{def:2}
(\textit{Task}).
\textnormal{An abstract service task $t$ within $w$ ($t \in ST$) is characterized by requirements for Confidentiality ($C$), Integrity ($I$), and Availability ($A$), as well as its Value to the workflow ($V$). Additionally, it is associated with a set of feasible Adaptation Actions, denoted by $\mathcal{AA}_t$, which encompasses a combination of actions from both Tenant Adaptation Actions and Middleware Adaptation Actions. Formally, $\mathcal{AA}_t = \{\alpha \alpha_t \mid \alpha \alpha_t \in \mathcal{T}\mathcal{AA} _{t} \cup \mathcal{M}\mathcal{AA} _{t}\}$.}
\end{definition}

\begin{example}\label{Example:1}
In Figure \ref{fig: WorkflowExample}, we illustrate the attributes ($C$, $I$, $A$, $V$, $\mathcal{AA}$) linked to each task in a running example.
\end{example}

\vspace{-5pt}
\begin{definition}\label{def:3}
(\textit{Tenant Adaptation Action}).
\normalfont Tenant Adaptation Actions ($\mathcal{T}\mathcal{AA}_{t_i}$) are actions at the tenant level within task $t_i$ to mitigate damage caused by detected violations. Each action is a tuple ($P$, $T$, $MI$, $V$) that represents the Price, Time, Mitigation Impact, and Value of the adaptation action for task $t_i$ in the workflow, respectively.
\end{definition}

\vspace{-5pt}
\begin{definition}\label{def:4}
(\textit{Middleware Adaptation Action}).
\normalfont Middleware Adaptation Actions ($\mathcal{M}\mathcal{AA}_{t_i}$) are actions at the middleware level within task $t_i$ to minimize damage caused by detected violations. Each action is a tuple ($P$, $T$, $MI$, $V$), indicating the Price, Time, Mitigation Impact, and Value of the adaptation action based on the parameters of task $t_i$ within the workflow, respectively.
\end{definition}

\vspace{-5pt}
\begin{definition}\label{def:6}
(\textit{Multi-Cloud Environment}).
\normalfont A multi-cloud environment comprises cloud services provided by different providers. Each service is described by a tuple ($P$, $T$, $C$, $I$, $A$, $AFR$), representing the Price, Time, Confidentiality, Integrity, Availability, and Attack Frequency Rate of each attack type within the service.
\end{definition}

\begin{table*}[h!]
\scriptsize
\fontsize{7pt}{8pt}\selectfont
\caption{{Attack Specifications based on \cite{yang2022network}} }
\label{table:Attack-Specification}
\def\arraystretch{1}
\ignorespaces 
\centering 
\input{tables/AttackSpecifications}

\end{table*}

\vspace{-5pt}
\begin{definition}\label{def:7}
(\textit{Multi-Tenant Environment}).
\normalfont A multi-tenant environment includes multiple Tenants, each defined by a tuple ($WS$, $Weights$,\linebreak $AdaptTriggerThresh$), indicating sets of workflows, preference weights, and trigger thresholds for reacting to attacks.
\end{definition}

\vspace{-30pt}
\begin{table*}[h!]
\vspace{10pt}
\scriptsize
\fontsize{7pt}{8pt}\selectfont
\caption{{The Properties of Different Adaptation Types} }
\label{table:Adaptation Types}
\def\arraystretch{1}
\ignorespaces 
\centering 
\input{tables/adaptationActions3}
\end{table*}




\vspace{-5pt}
\subsection{Problem Definition}
\vspace{-5pt}
In this section, we introduce a set of novel key terms, definitions, and algorithms to formally model the OACS problem.  To enhance the clarity of our approach, illustrative examples are provided for each definition.
\vspace{-5pt}
\begin{definition}\label{def:8}
 (\textit{Data Flow Closure Set}). 
\normalfont $DFCS_{t_i}$ signifies the set of tasks with direct or indirect data flow dependencies on task $t_i$. It is formally defined as:
\scriptsize
\begin{equation}
\begin{aligned}
DFCS_{t_i} = & \{ t_j \mid (t_i,t_j) \in E_d  \} 
& \cup \bigcup_{t_k \in DFCS_{t_i}} DFCS_{t_k}
\end{aligned}
\end{equation}
\end{definition}

\begin{example}\label{Example:2} the Data Flow Closure Set for task $t_5$ is denoted as $DFCS_{t_3}=\{t_5,t_6\}$. This set encompasses tasks that exhibit either direct or indirect data flow dependencies on task $t_3$.
\end{example}
\vspace{-5pt}
\begin{definition}\label{def:9}
(\textit{Control Flow Closure Set}).
\normalfont $CFCS_{t_i}$ represents the set of tasks holding direct or indirect control flow dependencies with task $t_i$. Its formal expression is given by:
\vspace{-5pt}
\scriptsize
\begin{equation}
\begin{aligned}
CFCS_{t_i} = & \{ t_j \mid (t_i, t_j) \in E_c \} 
& \cup \bigcup_{t_k \in CFCS_{t_i}} CFCS_{t_k}
\end{aligned}
\end{equation}
\end{definition}

\begin{example}\label{Example:3} In the provided scenario, the Control Flow Closure Set for task $t_3$ is represented as $CFCS_{t_3}=\{t_5,t_6,t_7,t_8\}$. This set encompasses tasks that exhibit either direct or indirect control flow dependencies with task $t_3$.
\end{example}
\vspace{-5pt}
\begin{definition}\label{def:10}
 (Security Dependency Matrix).
\normalfont The Security Dependency Matrix ($SDM$) for a workflow $w$, denoted as $SDM_w$, is a matrix that captures the relationships between tasks in terms of security. It quantifies the potential impact of attacks on the key security objectives, namely Confidentiality ($C$), Integrity ($I$),  and Availability ($A$), within the workflow. This matrix states the impact on the security properties of individual tasks and of adaptation actions when addressing detected violations. Hence it is used to evaluate and measure the impact of security violations and adaptation actions. The $SDM_w$ which is a $ST \times ST$ matrix, is presented in the following matrix format:
\vspace{-5pt}
\begin{equation}
\scriptsize
\label{eq:SDM_w}
  SDM_w=
    \begin{bNiceMatrix}[first-col,first-row]
        &t_1 &t_2 &\ldots &t_n \\
        t_1 & \text{\scriptsize ($C_{11}$,$I_{11}$, $A_{11}$)} &\text{\scriptsize ($C_{12}$,$I_{12}$, $A_{12}$)} & \ldots & \text{\scriptsize($C_{1n},I_{1n}, A_{1n}$)} \\
        t_2 &\text{\scriptsize ($C_{21},I_{21}, A_{21}$)}&\text{\scriptsize ($C_{22},I_{22}, A_{22}$)} &\ldots&\text{\scriptsize($C_{2n},I_{2n}, A_{2n}$)} \\
        \vdots & \vdots & \vdots & \ddots & \vdots \\
        t_n & \text{\scriptsize($C_{n1},I_{n1}, A_{n1}$)} & \text{\scriptsize($C_{n2},I_{n2}, A_{n2}$)} & \ldots & \text{\scriptsize($C_{nn},I_{nn}, A_{nn}$)}
    \end{bNiceMatrix}
\end{equation}

Here, each element [$t_i$, $t_j$] quantifies the security impact of security measures applied in task $t_i$ on task $t_j$. Essentially, it assesses how implementing security measures in one task affects the security objectives of other tasks in the workflow, providing insights into the cascading security effects within the workflow. 

To compute the Security Dependency Matrix ($SDM$), Algorithm \ref{alg:SecurityDependencyMatrix}  evaluates each task's impact on the Confidentiality ($C$), Integrity ($I$), and Availability ($A$) of other tasks within the workflow. By using the Data Flow Closure Set ($DFCS_{t_i}$) and Control Flow Closure Set ($CFCS_{t_i}$) from Definitions 8 and 9, respectively, the algorithm captures inter-task relationships in terms of security. For confidentiality, it examines backward and forward data flow connections between tasks $t_i$ and $t_j$ (lines 6-8), as a confidentiality breach in one task can affect tasks both preceding and succeeding the violated task. Regarding integrity, the algorithm focuses on forward data or control flow connections between $t_i$ and $t_j$ (lines 9-11). Specifically, it considers only forward data flow connections, as integrity violations do not propagate to tasks preceding the violated task. Additionally, control flow connections are included, as they may be influenced by integrity violations, such as decision-making tasks based on compromised data. Finally, for availability, the algorithm examines forward data flow connections between $t_i$ and $t_j$ (lines 12-14). Similar to integrity, availability violations do not propagate to tasks that precede the violated task.
\end{definition}
\vspace{-20pt}
\input{Algorithms/SDM_Calculation3}

\vspace{-20pt}

\begin{example}\label{Example:4} In the running example, Figure \ref{fig:SDMExample} displays the Security Dependency Matrix (SDM) corresponding to the workflow outlined in Figure \ref{fig: WorkflowExample}. This matrix is calculated using Algorithm \ref{alg:SecurityDependencyMatrix}.
\end{example}

\vspace{-35pt}
\begin{figure*}[htbp]
\begin{align*}
 \scriptsize
    \begin{bNiceMatrix}[first-col,first-row]
        & t_1 & t_3 & t_4 & t_5 & t_6 & t_8 \\
        t_1 & \text{\scriptsize 1, 1, 1} & \text{\scriptsize 0, 1, 0} & \text{\scriptsize 0, 1, 0} & \text{\scriptsize 0, 1, 0} & \text{\scriptsize 0, 1, 0} & \text{\scriptsize 0, 0.1, 0} \\
        t_3 & \text{\scriptsize 0, 0, 0} & \text{\scriptsize 1, 1, 1} & \text{\scriptsize 0, 0, 0} & \text{\scriptsize 0.25, 1, 1} & \text{\scriptsize 0.25, 1, 1} & \text{\scriptsize 0, 0.1, 0} \\
        t_4 & \text{\scriptsize 0, 0, 0} & \text{\scriptsize 0, 0, 0} & \text{\scriptsize 1, 1, 1} & \text{\scriptsize 0.35, 1, 0.5} & \text{\scriptsize 0.35, 1, 0.5} & \text{\scriptsize 0, 0.1, 0} \\
        t_5 & \text{\scriptsize 0, 0, 0} & \text{\scriptsize 0.25, 0, 0} & \text{\scriptsize 0.25, 0, 0} & \text{\scriptsize 1, 1, 1} & \text{\scriptsize 0.5, 1, 1} & \text{\scriptsize 0, 0.1, 0} \\
        t_6 & \text{\scriptsize 0, 0, 0} & \text{\scriptsize 0.5, 0, 0} & \text{\scriptsize 0.5, 0, 0} & \text{\scriptsize 1, 0, 0} & \text{\scriptsize 1, 1, 1} & \text{\scriptsize 0, 0.1, 0} \\
        t_8 & \text{\scriptsize 0, 0, 0} & \text{\scriptsize 0, 0, 0} & \text{\scriptsize 0, 0, 0} & \text{\scriptsize 0, 0, 0} & \text{\scriptsize 0, 0, 0} & \text{\scriptsize 1, 1, 1} \\
    \end{bNiceMatrix}
\end{align*}
\setlength{\belowcaptionskip}{-14pt}
\vspace{-20pt}
\caption{Security Dependency Matrix (SDM) for the running example}
\label{fig:SDMExample}
\end{figure*}

\begin{definition}\label{def:11}
(\textit{Security-Dependent Tasks}). 
\normalfont The set of Security-Dependent \linebreak Tasks ($DT_{vt}$), where $DT_{vt} \subseteq ST$, is defined as follows: \linebreak $DT_{vt}= \bigcup_{t_i \in ST} \{t_i \mid SDM[t_i, vt] \neq (0, 0, 0)\}$, where $SDM[t_i, vt]$ represents the Security Dependency Matrix for the task pair ($t_i$, $vt$), expressed as a triple ($C$, $I$, $A$). The set $DT_{vt}$ comprises tasks within the workflow that exhibit non-zero values in the Security Dependency Matrix, indicating an impact on the security objectives of the violated task $vt$.
\end{definition}

\begin{example}\label{Example:5} In our running example, the Security-Dependent Tasks for the task $t_5$ are represented as $DT_{t_5}=\{t_3, t_4, t_5, t_6, t_8\}$. This set encompasses tasks with non-zero values for at least one of the (C, I, A) components in the Security Dependency Matrix (depicted in Figure \ref{fig:SDMExample}).
\end{example}

\vspace{-5pt}
\begin{definition}\label{def:12}
(\textit{Adaptation Security Chain Set}).
\normalfont The Adaptation Chain Set ($\mathcal{AC}_{vt,a_k}$) encompasses all sets of adaptation chains designed to mitigate the impact of the detected attack $a_k$ in $vt$. Each adaptation chain $ac = (\alpha \alpha_{t_1}, \alpha \alpha_{t_2}, \ldots,$ $\quad$ $\alpha \alpha_{t_m})$ consists of adaptation actions $\alpha \alpha_{t_i}$, each associated with a specific task ($t_i$) within the workflow. This includes both backward and forward adaptations suitable for the attack $a_k$, encompassing adaptations to tasks positioned both preceding and succeeding the violated task and actions directly associated with the violated task itself. Therefore, $\mathcal{AC}_{vt,a_k}$ is defined as the Cartesian product of feasible actions contributing to mitigating the impact of the detected attack $a_k$, expressed as:

\begin{equation}
\scriptsize
 \mathcal{AC}_{vt,a_k} = \prod_{\forall t_i \in DT_{vt}} (\mathcal{AA}_{t_i} \cap MA_{\delta_{(a_k)}}) 
\label{eq:AdaptationChainSet}
\end{equation}

In this context, the intersection between the set of feasible adaptation actions for task $t_i$ ($\mathcal{AA}_{t_i}$) and the set of adaptation actions tailored to the severity ($\delta_{(a_k)}$) of the detected attack $a_k$ ($MA_{\delta_{(a_k)}}$) defines the set of feasible actions for each task. The process of generating the Adaptation Chain Set is defined by Algorithm \ref{alg:AdaptationChainSetGeneration}. The algorithm traverses each task $t_i$ in $DT_{vt}$ and, for every feasible adaptation action $aa_{t_i}$ in the intersection of $\mathcal{AA}{t_i}$ and $MA{\delta_{(a_k)}}$, it contributes to $\mathcal{AC}$ in two ways: a) it includes the current adaptation $(t_i,aa_{t_i})$ as an individual chain in the Adaptation Chain Set (line 4), b) it attaches the current adaptation $(t_i,aa_{t_i})$ to each existing chain in Adaptation Chain Set, creating a new chain if $t_i$ has not yet been associated with it (lines 5-8). Essentially, it ensures the exclusion of the same task from appearing more than once in a chain, preventing diverse adaptations within the same task in a single chain.
\end{definition}
\begin{example}\label{Example:6} 
Given a scenario involving a high-severity Denial of Service (DOS) attack on task $t_5$. Here, the Adaptation Security Chain Set 
is expressed as:
\vspace{-10pt}
\begin{equation*}
\scriptsize
\begin{aligned}
&\mathcal{AC}_{t_5,a_{DOS}} = \prod\biggl(t_3\rightarrow(\text{Insert},\text{Rework},\text{ReConfiguration}),\\
&\hspace{6.6em}t_4\rightarrow(\text{Insert},\text{Rework},\text{ReConfiguration}),\\
&\hspace{6.6em}t_5\rightarrow(\text{Insert},\text{Rework},\text{ReConfiguration}, \text{Redundancy}),\\
&\hspace{6.6em}t_6\rightarrow(\text{Insert},\text{Rework},\text{ReConfiguration}),\\
&\hspace{6.6em}t_8\rightarrow(\text{Insert},\text{Rework},\text{ReConfiguration})\biggr)
\end{aligned}
\vspace{-5pt}
\end{equation*}

Sample combinations from the set, outlining possible adaptation chains ($ac$) in response to the DOS attack on $t_5$, are:
\vspace{-5pt}
\input{Algorithms/chainExample3}

\end{example}
\vspace{-25pt}
\input{Algorithms/AdaptationSecurityChainSetGeneration3}

\vspace{-30pt}
\begin{definition}\label{def:13}
(\textit{Adaptation Chain Loops} ($\mathcal{ACL}_{vt,a_k}$)). 
\normalfont 
Such a loop encompasses all sets of adaptation chain loops ($acl$), each containing both intentional and unintentional adaptations. An ACL is called an adaptation process in the work \cite{reichert2009flexibility} and a process iteration body \cite{weiss2017model}.  Intentional adaptations refer to the adaptations specified within the adaptation chain $ac$, aimed at mitigating the impact of the detected attack $a_k$ in $vt$ (see Definition \ref{def:12}). However, due to the workflow's inherent data and control flow dependencies, these intentional adaptations ($ac$) may lead to unintended consequences, such as rework, appearing in tasks that have already been completed. This can occur when the starting points of the adaptation chain are positioned before the task $vt$, indicating that the chain's initiation lies within the set of tasks already completed. Consequently, any tasks completed between these starting points and the violated task must undergo rework due to the adaptations introduced by the adaptation chain. This is a problem known from runtime workflow adaptation research as reported in \cite{weiss2017model} We denote these unintended adaptations as $ac' = (\alpha \alpha_{t_1'}, \alpha \alpha_{t_2'}, \ldots, \alpha \alpha_{t_m'})$, which arise as a result of the adaptations within the adaptation chain $ac$ affecting tasks that were previously completed in the workflow instance.Thus, each adaptation chain loop ($acl = ac \text{\guillemotright} ac'$) encapsulates both intentional adaptation actions ($\alpha \alpha_{t_i}$) and unintentional adaptation actions ($\alpha \alpha_{t_i'}$). Within this algorithm, for each adaptation chain ($ac$) in $\mathcal{AC}$, the initial step involves identifying the starting points ($SPs$). To achieve this, the algorithm examines each task and adds it to the starting points if none of its predecessors are involved in the chain $ac$ (lines 3-5). Subsequently, for each starting point, the algorithm identifies the tasks that have already been completed (those situated between these starting points and task $vt$) and then incorporates these unintentional adaptations into the chain $ac$ (lines 6-9).
\end{definition}
\vspace{-10pt}
\begin{example}\label{ch5:Example:7} 
\normalfont In Example \ref{Example:6}, there are no intentional adaptations (\textit{ac}) that lead to unintended consequences, as no tasks exist between the starting point of the chain ($t_3$) and the violated task ($t_5$). However, if we consider the hypothetical adaptation 
$ac_x: \{(\text{Insert}_{t_1}) \text{\guillemotright} (\text{Rework}_{t_5})\},$
the execution of $ac_x$ would result in unintentional adaptations involving tasks $t_3$ and $t_4$, as these tasks lie between the starting point of the adaptation chain ($t_1$) and the end of the adaptation chain ($t_5$). Consequently, the adaptation chain $x$ would be modified as follows:
$acl_x: \{(\text{Insert}_{t_1})\text{\guillemotright} (\text{Rework}_{t_3}) \text{\guillemotright} (\text{Rework}_{t_4}) \text{\guillemotright} (\text{Rework}_{t_5})\}.$
\end{example}
\vspace{-25pt}
\input{Algorithms/AdaptationChainLoopGeneration}

\vspace{-25pt}

\begin{definition}\label{def:14}
(\textit{Adaptation Chain Constraints}).
\normalfont These constraints outline the limitations associated with simultaneously implementing two or more adaptations within a single workflow instance.  As part of the modeling phase, tenants extract and seamlessly integrate these constraints into the workflow model. We categorize these constraints into two types: a) Conflicting Dependency: These constraints address conflicts that may arise between actions in an adaptation chain. They are represented as $aa_{t_i} \triangle aa_{t_j}$, indicating that if adaptation $aa_{t_i}$ is performed in task $t_i$ within a single workflow instance, the adaptation $aa_{t_j}$ in task $t_j$ cannot occur simultaneously; b) Essential Dependency Constraints: These constraints signify the necessity between actions in an adaptation chain. Represented as $aa_{t_i} \vdash aa_{t_j}$, they convey that if adaptation $aa_{t_i}$ is applied to task $t_i$ within a single workflow instance, the adaptation $aa_{t_j}$ in task $t_j$ should also be performed.

Each constraint defines as: $const\subseteq \bigcup_{t \in ST,aa_t \in \mathcal{AA}} aa_t$. By applying these constraints, we can eliminate adaptation chains containing actions constrained by these conditions. Algorithm \ref{alg:ChainConstraintResolver} demonstrates how these constraints are applied and the subsequent creation of Resolved Adaptation Chains ($\mathcal{RAC}$).
\end{definition}

\begin{example}\label{Example:8} In our ongoing scenario, we introduce three adaptation chain constraints as follows:
\begin{equation*}
\scriptsize
\begin{aligned}
&const_1 : (\text{Insert}_{t_4}) \triangle  (\text{Insert}_{t_5})\\
&const_2 : (\text{Skip}_{t_3}) \triangle (\text{Insert}_{t_4})\\
&const_3 : (\text{Late}_{t_4}) \vdash (\text{Insert}_{t_5})\\
\end{aligned}
\end{equation*}
$const_1$ specifies that simultaneously adapting tasks $t_4$ and $t_5$ using the "Insert" type within a single workflow instance is prohibited.
$const_2$ dictates that adaptations involving "Skip" in task $t_3$ and "Insert" in task $t_4$ cannot take place at the same time. Additionally, $const_3$ introduces a dependency, stating that if the adaptation with the "Late" type in task $t_4$ is applied, then task $t_5$ must also undergo adaptation with the "Insert" type. Consequently, based on the $const_1$, the adaptation chain $ac_4$ in Example \ref{Example:6} will be removed from the set $\mathcal{RAC}$.
\end{example}
\vspace{-25pt}
\input{Algorithms/ChainConstraintResolver}

\vspace{-25pt}
\begin{definition}\label{def:15}
 \textit{Adaptation Chain Cost} ($\phi_{ac}$).
\normalfont 
This corresponds to the total cost associated with the adaptation chain $ac$, intended to mitigate the impact of the detected attack on the violated task $vt$, taking into account the previously selected adaptation chains in $\mathcal{ACL}_{history}$. More precisely, the cost of the adaptation chain $ac$ is determined while considering the adaptations already made in the workflow instance. The overall time, price, value, and mitigation score of the adaptation chain $ac$ are computed using the following formula:

\begin{equation}
\scriptsize
\phi_{ac} = \sum_{att_i \in \{P,T, V, MS\}} W_{att_i} \cdot att_i(w)
\label{eq:AdaptationChainCost}
\end{equation}

where $att_i$ represents the observed values for P (price), T (time), V (value), and MS (mitigation score) for the entire workflow after applying the adaptation chain $ac$. The weighting factor $W_{att_i}$ is positive for price and time and negative for mitigation score and value. 

For each adaptation action within the chain, the time, price, and value are directly assigned based on the adaptation type in the associated task (see Table \ref{table:Adaptation Types}). The mitigation score is calculated using Equation \ref{Equation:MitigationScore}. This equation considers the security requirements of task $t_i$ (represented by $obj_{t_i}$), the impact of the detected attack $a_k$ on the CIA aspects (represented by $obj_{a_k}$), the mitigation impact of the adaptation action on each aspect (represented by $obj_{MI_{aa}}$), and the Security Dependency between $t_i$ and $vt$ (represented by $obj_{SDM[vt][t_i]}$).

\begin{multline}
\centering
MS_{aa_{t_i},vt,a_k} = 
\sum_{obj \in {C,I,A}} (1 - obj_{t_i} \cdot obj_{a_k}) \cdot obj_{MI_{aa}}\cdot obj_{SDM[vt][t_i]}
\label{Equation:MitigationScore}
\end{multline}

Algorithm \ref{alg:AdaptationChainCostCalculation} outlines how the Adaptation Chain Cost is computed for each adaptation chain ($ac$) in $\mathcal{RAC}$. 
Firstly, it checks if introducing the current chain ($ac$) conflicts with any Adaptation Chain Constraints (Definition \ref{def:14}), setting the total cost of $ac$ to infinity if any conflicts are found (lines 2-3). Then, it iterates over tasks positioned between the successors of the previously violated task and the predecessors of the current violated task, calculating costs for any postponed adaptations or normal task executions (lines 4-9).

Next, the algorithm computes the cost of the current adaptation chain ($ac$) by evaluating the contribution of each task (lines 10-18). For each task in the predecessors of the violated task, it checks for the required adaptation action ($aa_{current}$) and adds its associated costs if present (lines 12-14). 
\vspace{-20pt}
\input{Algorithms/AdaptationChainCostCalculation7}
\vspace{-20pt}
In the case of the violated task, if multiple options exist, the algorithm selects the current chain's option and removes all other adaptations from the history (line 15). If $aa_{current}$ is empty, it adds the cost of the last adaptation from history ($aa_{history}$) (line 17). If both $aa_{current}$ and $AA_{history}$ are empty, the algorithm includes the default task cost (line 18). Finally, the total cost of $ac$ is calculated using Equation \ref{eq:AdaptationChainCost} (line 19).

\end{definition}



\begin{definition}\label{def:16}
(\textit{Optimal Adaptation Chain Selection Problem}).
\normalfont The Optimal Adaptation Chain Selection (OACS) Problem has as its primary goal to determine the optimal adaptation chain ($ac^*$) for a detected attack $a_k$ within the task $vt$. 
Therefore, for a given $vt$, $a_k$, and the adaptation action set $\mathcal{ACL}_{vt,a_k}$ for the workflow $w$, where each adaptation chain ($ac \in \mathcal{ACL}_{vt,a_k}$) incurs a cost denoted by ($\phi_{ac}$), the primary goal is to identify the adaptation chain that minimizes the cumulative Adaptation Chain Cost:

\begin{equation}
\scriptsize
ac^* = \arg\min_{ac \in \mathcal{ACL}_{vt,a_k}} \phi_{ac}
\label{eq:optimal_chain}
\end{equation}

Here, $ac^*$ represents the optimal adaptation chain, and $\arg\min$ denotes the adaptation chain within $\mathcal{ACL}_{vt,a_k}$ that achieves the minimum Adaptation Chain Cost ($\phi_{ac}$).

After identifying the Optimal Adaptation Chain ($ac^*$), Algorithm \ref{alg:ApplySelectedAdaptationChain} illustrates how to adapt the current workflow instance ($current_{inst}$) by following this selected chain during execution. In this algorithm, the initial step involves suspending the $current_{inst}$ (line 1), followed by identifying the starting point of the adaptation chain (lines 2-4). Subsequently, the $current_{inst}$ is resumed from the specified starting point to execute tasks according to the planned adaptations outlined in the chain $ac^*$ (lines 5-12).
\vspace{-23pt}
\input{Algorithms/ApplySelectedAdaptationChain2}
\vspace{-30pt}
\end{definition}

%% file: tables/AttackSpecifications.tex
\vspace*{-.4cm}
\begin{tabulary}{\linewidth}{p{1cm}p{2cm}p{2.2cm}p{3cm}p{3.5cm}}
 \hline
 \textbf{} & \textbf{Impact on} &   \multicolumn{2}{c}{\textbf{Mitigation Actions}} & \textbf{}\\ 

  \textbf{AT} &
 \textbf{ (C,I,A)} &

\textbf{(Low,} &
\textbf{Medium,}&
\textbf{High)} 
\\  \hline  \hline
\textbf{DoS} &
(0.56,0.56,0.56)	&Switch, Rework	&Insert, Rework 	&Insert, Rework, Redundancy, ReConfiguration
\\  \hline

\textbf{Probe} &
(0.22,0.22,0)&	Skip&	 Skip, ReConfiguration& 	Skip, ReConfiguration
\\  \hline

 \textbf{U2R} &
(0.56,0.22,0.22)	&Insert, Rework 	&  Insert, Rework	& Insert, Rework, Redundancy, ReConfiguration
\\  \hline

 \textbf{R2L} &
0.56,0.56,0.22)&	Rework&	Insert, Rework &	 Insert, Rework, ReConfiguration
\\  \hline

\bottomrule 
\end{tabulary}\par 
\vspace*{-.4cm}

%% file: tables/adaptationActions3.tex
\vspace*{-.4cm}
\begin{tabulary}{\linewidth}{ 
p{2.5cm} p{2.8cm} p{2cm} p{2.3cm} p{2cm}}
 \hline
 \textbf{AdaptType} &
 \textbf{T}  &
\textbf{P}  & 
\textbf{V}  &
\textbf{MI(C,I,A)}
\\  \hline  \hline

\textbf{Insert} &
$T_{newTask}$ & $P_{newTask}$ & $V_{newTask}$ &  $(0.7,0.9,0.9)$
\\  \hline

\textbf{Switch} &
$T_{Switch}$ & $P$  &  $V_{Switch}$ & $(0.7,0.6,0.8)$
\\  \hline

\textbf{Skip} &
$0$  & $0$ &  $0$ & $(0.5,0.4,0.6)$ 
\\  \hline

\textbf{Rework} &
$T_{BackupSrc}$  & $P_{BackupSrc}$ & $V$ & $(0.5,0.9,0.7)$
\\  \hline

\textbf{Redundancy} &
 $Max(T_{BackupSrc},T)$ & $P+P_{BackupSrc}$ & $V+V_{Redundancy}$ & $(0.5,0.8,0.9)$
\\  \hline

\textbf{Reconfiguration} &
$T+T_{reconfig}$ & $P+P_{reconfig}$ & $V+V_{Reconfig}$ & $(0.6,0.7,0.5)$
\\  \hline

\bottomrule 
\end{tabulary}\par 
\vspace*{-.4cm}

%% file: Algorithms/SDM_Calculation3.tex
\begin{algorithm}
\scriptsize
\caption{Security Dependency Matrix Computation}
\label{alg:SecurityDependencyMatrix}

\SetKwInOut{Input}{Input}
\SetKwInOut{Output}{Output}

\Input{$w$:($ST$, $D$, $E_c$, $E_d$)}
\Output{$SDM_w$}

\BlankLine

\ForEach{$t_i$ in $ST$}{
    \ForEach{$t_j$ in $ST$}{
        \If{$t_i \neq t_j$}{
            \If{$t_j \in DFCS(t_i) \,||\, t_i \in DFCS(t_j) $}{
                $C_{ij} \gets \prod_{\substack{t_k \\ (t_i \leq t_k \leq t_j) \,||\, (t_j \leq t_k \leq t_i)}} C_{t_k}$\;
            }
          
            \If{$t_j \in \text{DFCS}(t_i) \,||\, t_j \in \text{CFCS}(t_i) $}{
                $I_{ij} \gets \prod_{\substack{t_k \\ t_i \leq t_k \leq t_j}} I_{t_k}$\;
            }
            \If{$t_j \in DFCS(t_i) $}{
                $A_{ij} \gets \prod_{\substack{t_k \\ t_i \leq t_k \leq t_j}} A_{t_k}$\;
            }
        }\Else{
            $C_{ij} \gets 1, I_{ij} \gets 1, A_{ij} \gets 1$\;
        }
        $SDM_w[t_i][t_j] \gets (C_{ij}, I_{ij}, A_{ij})$\;
    }
}
\end{algorithm}

%% file: Algorithms/chainExample3.tex
\begin{equation*}
\scriptsize
\begin{aligned}
&ac_1: \{(\text{Insert}_{t_5})\}\\
&ac_2: \{(\text{Insert}_{t_1})  \text{\guillemotright} (\text{Rework}_{t_5})\}\\
&ac_3:\{(\text{Insert}_{t_3})  \text{\guillemotright} (\text{ReConfiguration}_{t_5})\text{\guillemotright}(\text{Rework}_{t_6})\}\\
&ac_4: \{( \text{Insert}_{t_1})  \text{\guillemotright} (\text{Rework}_{t_3}) \text{\guillemotright}(\text{Insert}_{t_4})\text{\guillemotright}(\text{Insert}_{t_5}) \text{\guillemotright}(  \text{Rework}_{t_6})\}
\end{aligned}
\end{equation*}

%% file: Algorithms/AdaptationSecurityChainSetGeneration3.tex
\begin{algorithm}
\scriptsize
    \caption{Adaptation Chain Set Generation}
    \label{alg:AdaptationChainSetGeneration}
    \SetKwInOut{Input}{Input}
    \SetKwInOut{Output}{Output}

   \Input{
    $\mathcal{AA}_{t_i}$, $MA_{\delta_{(a_k)}}$, $DT_{vt}$
    }
    \Output{
        $\mathcal{AC}_{vt,a_k}$
    }
    \BlankLine
    $\mathcal{AC} \gets \emptyset$;\\
    \ForEach{$t_i$ in $DT_{vt}$}{
        \ForEach{$aa_{t_i}$ in $(\mathcal{AA}_{t_i} \cap MA_{\delta_{(a_k)}})$}{
            $\mathcal{AC}_{new} \gets \mathcal{AC} \cup \{(aa_{t_i})\}$\;
            \ForEach{$ac$ in $\mathcal{AC}$}{
              \If{$t_i \notin \{t \mid aa_t \in ac\}$}{
                    $ac \gets ac \text{\guillemotright}(aa_{t_i})$\;
                    $\mathcal{AC}_{new} \gets \mathcal{AC}_{new} \cup \{ac\}$\;
                }
            }
            $\mathcal{AC} \gets \mathcal{AC}_{new}$;\\
        }
    }
\end{algorithm}

%% file: Algorithms/AdaptationChainLoopGeneration.tex
\begin{algorithm}
\scriptsize
    \caption{Adaptation Chain Loop Generation}
    \label{alg:AdaptationChainLoopGeneration}
    \SetKwInOut{Input}{Input}
    \SetKwInOut{Output}{Output}

   \Input{
    $\mathcal{AC}_{vt,a_k}$
    }
    \Output{
        $\mathcal{ACL}_{vt,a_k}$
    }
    \BlankLine
    $\mathcal{ACL} \gets \emptyset$;\\
    \ForEach{$ac$ in $\mathcal{AC}$}{
        \ForEach{$t_i$ in $ac$}{
            \textbf{if }$pred_{t_i} \cap \{t \mid aa_t \in ac\}= \emptyset$ \textbf{ then } 
                 $SPs \gets SPs \cup t_i$;\\
           
        }
        \ForEach{$sp$ in $SPs$}{
            \ForEach{$t_i$ in $ST$}{
                \textbf{if }$t_i \in pred_{vt} \land t_i \in succ_{sp} \land t_i \notin ac$ \textbf{ then } 
                    $ac \gets ac \text{\guillemotright}(Rework_{t_i})$\;
                
            }
        }
        $\mathcal{ACL} \gets \mathcal{ACL} \cup \{ac\}$\;
    }
\end{algorithm}

%% file: Algorithms/ChainConstraintResolver.tex
\begin{algorithm}
\scriptsize
\caption{Chain Constraint Resolver }
\label{alg:ChainConstraintResolver}

\SetKwInOut{Input}{Input}
\SetKwInOut{Output}{Output}

\Input{$ \mathcal{A}\mathcal{CL}_{vt,a_k}$}
\Output{$ \mathcal{RAC}_{vt,a_k}$}

\BlankLine
$\mathcal{RA}\mathcal{C}\gets\mathcal{A}\mathcal{CL}$;

\ForEach{$ac$ in $\mathcal{A}\mathcal{CL}$}{
    \ForEach{$const$ in Constraints}{
        \textbf{if }$const$  matches   $ac$\textbf{ then } 
           Remove $ac$ from $\mathcal{RA}\mathcal{C}$;
    }
}
\end{algorithm}

%% file: Algorithms/AdaptationChainCostCalculation7.tex
\begin{algorithm}
\scriptsize
\caption{Adaptation Chain Cost Calculation}
\label{alg:AdaptationChainCostCalculation}

\SetKwInOut{Input}{Input}
\SetKwInOut{Output}{Output}

\Input{$w,\mathcal{RAC}_{vt,a_k},\mathcal{ACL}_{history},W_{P},W_{T},W_{V},W_{MS} $}
\Output{$\forall ac \in \mathcal{RAC}: \phi_{ac}$}

\BlankLine
\ForEach{$ac$ in $\mathcal{RAC}$}{
    \If{$\exists const \in Constraints$ such that $const \text{ matches } (ac \cup AA_{history})$}{
        $\phi_{ac}=\infty$;
    }
 
        \ForEach{$t_i$ in  ($succ_{previous_{vt}}\cap pred_{vt})$ }{
             $aa_{history} \gets \emptyset$;\\
             \If{$ t_i \in  \{t \mid \exists ac_h \in \mathcal{AC}_{history}, aa_t \in ac_h\}$}{
                    $aa_{history} \gets aa_{t_i}$ from the last applied $ac$;\\
            } 
                \textbf{if }$aa_{history} \neq \emptyset $ \textbf{ then }  
                $ac_{P,T,V,MS} \mathrel{+}= aa_{{history}_{P, T, V, MS}}$;\\
            \textbf{if } $AA_{history} == \emptyset$ \textbf{ then } 
                $ac_{P,T,V,MS} \mathrel{+}= Default_{P, T, V, MS}$;\\
             
        }
        \ForEach{$t_i$ in $(vt \cup pred_{vt})$}{
        $aa_{current} \gets \emptyset$;\\
            \textbf{if }$t_i \in \{t \mid aa_t \in ac\}$\textbf{ then }
                $aa_{current} \gets aa_{t_i}$;\\

            \If{$aa_{current} \neq \emptyset$}{
            $ac_{P,T,V,MS} \mathrel{+}= aa_{{current}_{P, T, V, MS}}$;\\
                  \textbf{if }$t_i == vt$\textbf{ then }
                    Remove all other $aa_{vt}$ from $\mathcal{AC}_{history}$
            }
            \If{$t_i == vt \land aa_{current} == \emptyset $  }{      
                \textbf{if }$aa_{history} \neq \emptyset$\textbf{ then }
                $ac_{P,T,V,MS} \mathrel{+}= aa_{{history}_{P, T, V, MS}}$;\\
                \textbf{else }
                    $ac_{P,T,V,MS} \mathrel{+}= Default_{P, T, V, MS}$;\\
            }
        }
    }
    $\phi_{ac}=W_{P} \cdot ac_{P}+ W_{T} \cdot ac_{T} +W_{V} \cdot ac_{V} +W_{MS}\cdot ac_{MS}$;
\end{algorithm}

%% file: Algorithms/ApplySelectedAdaptationChain2.tex
\begin{algorithm}
\scriptsize
\caption{Applying Selected Adaptation Chain}
\label{alg:ApplySelectedAdaptationChain}

\SetKwInOut{Input}{Input}
\SetKwInOut{Output}{Output}

\Input{$w:(ST, D, E_c, E_d),ac^*, current_{inst}$}
\Output{Applying $ac^*$ within $current_{inst}$}

\BlankLine
Suspend $current_{inst}$;\\
$t_{start}\gets vt$;\\
\ForEach{$t_i$ in $ pred_{vt}$}{
    \textbf{if }$t_i \in \{t \mid aa_t \in ac\}$ \textbf{ then } $t_{start}=t_i$ 
}
Resume $current_{inst}$ from $t_{start}$;\\
\ForEach{$t_j$ in $(t_{start} \cup succ_{t_{start}} )$}{
    $aa_{current} \gets \emptyset$;\\
    \If{$t_j \in \{t \mid aa_t \in ac\}$}{
        $aa_{current} \gets aa_{t_j}$;\\
        Adapt $t_j$ with $aa_{current}$;\\
    }
    \Else{
        Repeat $t_j$;\\
    }
}

\end{algorithm}

%% file: Sections/4_ProposedMethod.tex
\section{Reinforcement Learning Based Solution for the  OACS Problem}
\vspace{-10pt}
Addressing the OACS Problem described in Definition \ref{def:16}, we propose employing RL to identify the optimal adaptation chain under consideration of previous reactions to mitigate the uncertain costs associated with adaptations. These costs are linked to various factors, including the current workflow state, previous violations, and workflow complexity, all of which significantly influence the risks and costs of task adaptations \cite{soveizi2023enhancing}. By employing RL, this tailored adaptation chain addresses these uncertain costs by factoring in attack characteristics, the inherent nature of the workflow, and user-defined requirements. Furthermore, unlike a single, fixed adaptation, the adaptation chain offers increased flexibility by permitting the selection of multiple adaptation actions, ensuring sufficient mitigation of attack impact, optimized trade-offs, tailored customization, and effective handling of complex attack scenarios. To model the problem, we employ the Markov Decision Process ($MDP$), which we will describe in detail in this section.

A $MDP$ is a 4-tuple $MDP$=($State$, $Action$,  $Probability$, $Reward$), with the following definitions: 1) \textbf{$State$}: Represents all possible states. Each state, denoted by a 2-tuple $st=(St_T, St_W)$, comprises the current task state $St_T$, including the detected attack ($a_k$) and its severity ($\gamma_{(a_k)}$), and the workflow state $St_W$, containing information on past violations, adaptation chains, and the workflow's current status. 2)\textbf{$Action$}: Denotes the potential adaptation chains available at a state $st$. The action set $A(st)$ includes all adaptation chains that can be initiated at $st$, expressed as $A(st) = \mathcal{RA}\mathcal{C}_{vt,a_k}$ (See Algorithm \ref{alg:ChainConstraintResolver}). 3)\textbf{$Probability$}: Describes the likelihood of transitioning between states when executing a specific action. It is denoted by the probability distribution $P(st'|st, a)$. 4) \textbf{$Reward$}: Measures the effectiveness of adaptation chain selection. If an adaptation chain \(ac\) is chosen, the reward function is defined as the Adaptation Chain Cost: \(R(ac) = \phi_{ac}\) (See Algorithm \ref{alg:AdaptationChainCostCalculation}).


The mean Q-value of action $a$ on state $st$ following policy $\pi$ is denoted as $Q_{\pi}(st, a)$. The optimal Q-value function is defined as:
{\scriptsize
\begin{align}
Q(st,a) = \sum_{st'} \gamma(st'|st,a) \left[R(st' | st, a) + \gamma \min_{a'} Q(st',a')\right]
\end{align}}

Here, $\gamma$ represents the discount factor, $R(st'|st, a)$ is the reward received when transitioning from state $st$ to $st'$ by performing action $a$, and $\min_{a'} Q^*(st',a')$ calculates the minimum Q-value for the next state $st'$. This optimal value function is nested within the Bellman optimality equation (refer to Algorithm \ref{alg:TrainingActionSelectionAlgorithm}).

Algorithm \ref{alg:WholeAlgorithm} presents the whole procedure for adapting the workflow based on the chosen adaptation chain. If any violations occur during the workflow execution (line 5), the optimal chain will be selected based on Algorithm \ref{alg:TrainingActionSelectionAlgorithm} (line 6), and subsequently implemented using Algorithm \ref{alg:ApplySelectedAdaptationChain} (line 7).

\vspace{-20pt}
\input{Algorithms/RL2}
\vspace{-25pt}
\vspace{-20pt}
\input{Algorithms/WholeProcedure}
\vspace{-12pt}
\vspace{-10pt}

\begin{example}\label{ch5:Example:9}
\normalfont In the provided scenario, based on Example \ref{Example:6}, each task in the chain contributes to the adaptation process in response to the DOS attack on $t_5$, offering multiple options. For instance, $t_3$ has three primary adaptation actions—Insert, Rework, and Reconfiguration—along with an implicit option where the task does not contribute to the adaptation chain. Considering all tasks involved ($t_3, t_4, t_5, t_6, t_8$), the total number of possible adaptation chains can be calculated as: 4×4×5×4×4=1280.

This calculation accounts for four possible states for each task, except $t_5$, which has five states due to the inclusion of Redundancy. Table~\ref{tab:chain_costs} details the price, time, value, mitigation score, and total cost for the alternative chain examples introduced in Example \ref{Example:6} ($ac_1$ to $ac_4$). Furthermore, Table~\ref{tab:chain_costs} highlights the best-identified chain, which achieved the lowest total cost during a high-severity DOS attack, as determined using the proposed method, among all 1280 possible chains.

\begin{table}[ht]
\centering
\caption{Examples of Adaptation Chains Evaluated for the Provided Scenario}
\label{tab:chain_costs}
\begin{tabular}{|c|c|c|c|c|c|}
\hline
\textbf{Chain} & \textbf{Time (s)} & \textbf{Price} & \textbf{Value} & \textbf{Mitigation Score} & \textbf{Total Cost} \\ \hline
Best Chain     & 38.5              & 21.14          & 65.0           & 1.36                      & 1.74                \\ \hline
$ac_1$         & 19.0              & 12.04          & 15.0           & 0.31                      & 4.78                \\ \hline
$ac_2$         & 70.0              & 8.55           & 45.0           & 1.59                      & 9.90                \\ \hline
$ac_3$         & 77.0              & 24.99          & 55.0           & 2.21                      & 13.87               \\ \hline
$ac_4$         & 161.0             & 58.80          & 50.0           & 1.61                      & 50.78               \\ \hline
\end{tabular}
\end{table}

The following sequence of actions represents the best-identified chain:

\begin{equation*}
\begin{aligned}
&ac_{best}: \{( \text{Redundancy}_{t_5})  \text{\guillemotright} (\text{Insert}_{t_6})\}
\end{aligned}
\end{equation*}

\end{example}

%% file: Algorithms/RL2.tex
\begin{algorithm}
\scriptsize
\caption{Action Selection Trainer Algorithm}
\label{alg:TrainingActionSelectionAlgorithm}

\KwIn{ $st$-The current state}
\KwOut{Trained Model for Selecting the proper Adaptation Chain}
Calculate $SDM$ for $w$ based on \ref{alg:SecurityDependencyMatrix};\\
\For{each episode}{
  $st \gets st_0$ \\
  \For{$st \notin St_r $}{
    Generate $\mathcal{RAC}$ based on \ref{alg:AdaptationChainSetGeneration}, \ref{alg:AdaptationChainLoopGeneration},\ref{alg:ChainConstraintResolver};\\
    Choose $ac \in \mathcal{RAC}$ based on $\epsilon$-greedy policy\;
    Perform $ac$, observe reward $r$ and new state $s'$ based on \ref{alg:AdaptationChainCostCalculation}\;
    $Q(st,ac) \gets Q(st,ac)+ \alpha \left[r + \gamma \min_{a'} Q(st',ac')- Q(st,ac)\right]$\;
    $st \gets s'$\;
  }
}
\end{algorithm}

%% file: Algorithms/WholeProcedure.tex
\begin{algorithm}
\scriptsize
\caption{Real-time Workflow Monitoring and Adaptation}
\label{alg:WholeAlgorithm}

\SetKwInOut{Input}{Input}
\SetKwInOut{Output}{Output}

\Input{$w:(ST, D, E_c, E_d)$}
\Output{Real-time Workflow Monitoring and Adaptation}

\BlankLine
\While{system is operational}{
    \ForEach{$t_i$ in $ST$}{
        \While{Execute($t_i$)}{
        Monitor Services based on \cite{soveizi2023enhancing};\\
          \If {any Violation is detected in $t_i$}{
                Select $ac^*$ based on trained model in \ref{alg:TrainingActionSelectionAlgorithm};\\
                Apply $ac^*$ based on \ref{alg:ApplySelectedAdaptationChain};\\
           }
        }
 
    }
}
\end{algorithm}

%% file: Sections/5_Evaluation.tex
 \section{Evaluation}
 \label{sec:Evaluation}
 \vspace{-10pt}
We implemented our method by extending the jBPM (Java Business Process Management) \cite{jBPM} engine and integrating it with the Cloudsim Plus \cite{cloudsimplus} simulation tool. 
To evaluate our approach, we utilized process models comprising 10 to 50 tasks. Our scenario assumed the availability of 5 cloud providers, each offering 3 different services for the service tasks. The specifications of these services fell within the following ranges: Response time [1, 50], Cost [0.1, 10], and confidentiality, integrity, and availability [0, 1]. The response times are selected randomly such that the fastest service is roughly three times faster than the slowest one, and accordingly, it is roughly three times more expensive. Furthermore, in Section \ref{sec:Problem Model}, Table \ref{table:Adaptation Types} offers a summary of the adaptation type properties used in our experiments, specifically,  Denial of Service (DoS), probe attacks, Remote-to-Local (R2L), and User-to-Root (U2R), as outlined in Table \ref{table:Attack-Specification}. 

\begin{figure}[H]
\vspace{-12pt}
  \centering
  \includegraphics[scale=0.5]{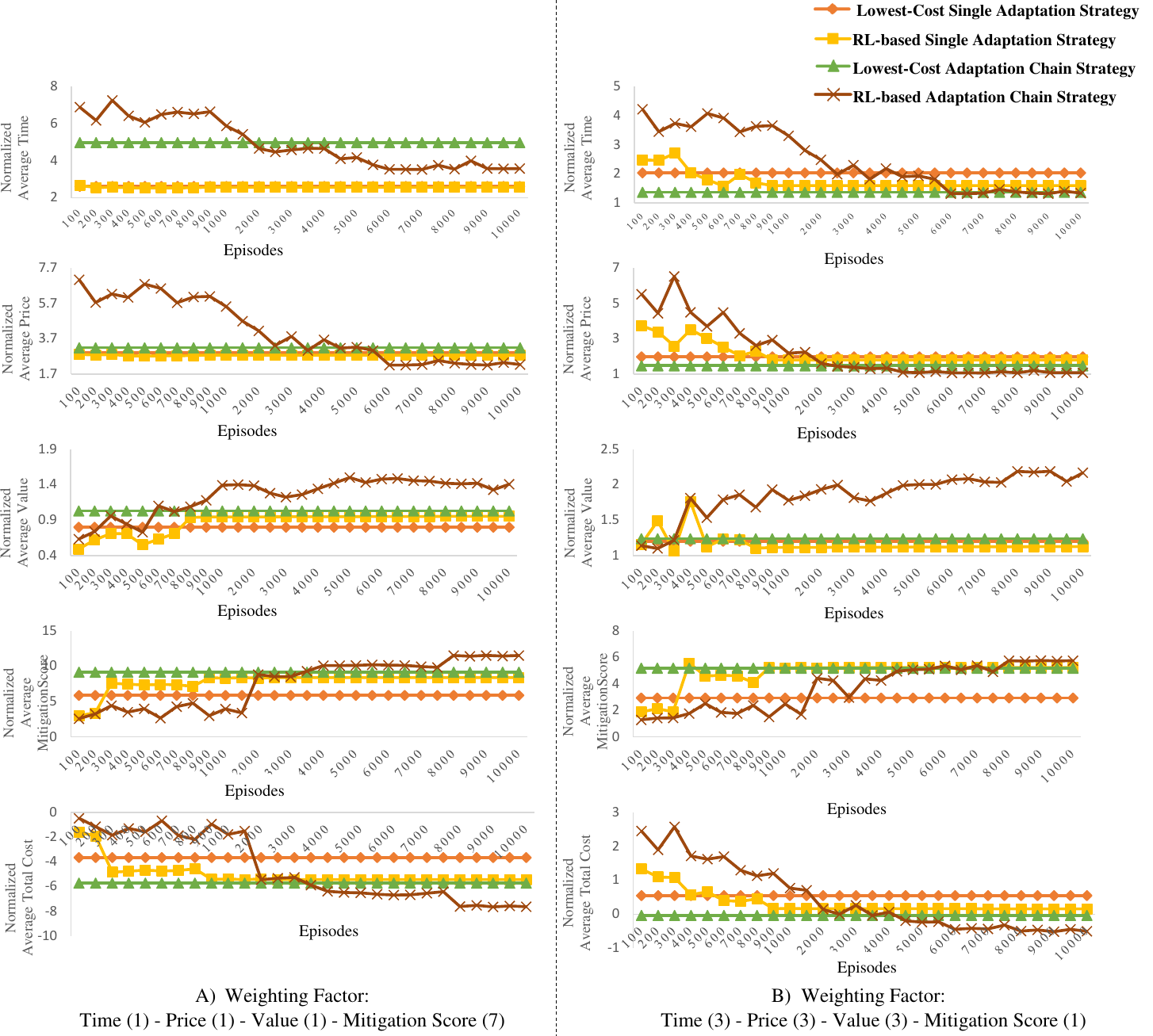}
  \vspace{-20pt}
  \caption{Comparison of Metrics for Various Adaptation Strategies.}   
  \label{fig:large}
  \vspace{-20pt}
\end{figure}

 We conducted 10,000 executions at an attack rate of 0.3, analyzing the average time, price, value, and mitigation score for each strategy. Results are presented for two distinct weighting factors: 1-1-1-7 and 3-3-3-1, representing varied priorities for time, price, value, and mitigation score. Notably, for the RL-based Strategies (both single and chain), averages were computed across every 1000 executions. Figure \ref{fig:large} illustrates the time, price, value, mitigation score, and total cost ($\phi_{\text{total}}$ based on Equation \ref{eq:AdaptationChainCost}), with subfigures A and B corresponding to Weighting Factors 1-1-1-7 and 3-3-3-1, respectively. Additionally, it's worth mentioning that we conducted experiments for the small class (5-10 tasks) as well, although the figures are not shown due to space constraints; however, the results exhibit a consistent trend. Our findings underscore the effectiveness of the RL-based Adaptation Chain Strategy compared to the RL-based Single Adaptation Strategy, particularly in reducing total costs. We also observe that the RL-based Adaptation Chain Strategy necessitated more time to identify the optimal set compared to the RL-based Single Adaptation Strategy. 

 This can be attributed to the broader solution space entailed in the adaptation chain approach. Furthermore, as expected, when utilizing weighting factors of 1-1-1-7, the algorithm tended to prioritize adaptations with the highest mitigation scores. Conversely, in scenarios characterized by weighting factors of 3-3-3-1, the algorithm demonstrated a preference for selecting adaptations that optimized cost and time efficiency. 


%% file: Sections/6-related2.tex
\vspace{-10pt}
\section{Related Works}
 \vspace{-10pt}
\label{Related Work}
 Based on a recent systematic review encompassing both business and scientific workflows' security and privacy aspects \cite{soveizi2023security}, we conclude that across the relevant works, a notable limitation is the absence of a structured solution for adaptation as a reaction to security. Existing approaches often focus on singular reactions to detected violations, which may not sufficiently mitigate risks associated with various violation types. Only one approach \cite{soveizi2023enhancing} systematically addresses potential attacks compromising the Confidentiality, Integrity, and Availability (CIA) of outsourced workflow tasks in multi-cloud environments. However, this work \cite{soveizi2023enhancing} primarily relies on single adaptation actions, leading to inadequate mitigation of attack impact. This limitation stems from not considering control and data dependencies between tasks, an inability to address conflicting objectives in scenarios with multiple goals, and vulnerability to delays in complex attack scenarios.

%

%% file: Sections/7_Conclusion.tex
 \vspace{-10pt}
\section{Conclusion}
 \vspace{-10pt}
\label{sec:Conclusion}
In this paper, we've tackled significant research gaps concerning responses to security violations in cloud-based workflow execution. We introduced an RL-based adaptive strategy to select the optimal workflow adaptation chain for each violation. Our approach considers factors such as the attack's characteristics, the workflow's nature, and user-defined requirements, tailoring our response accordingly. Unlike single-task adaptation constraints, adaptation chains we introduced here ensure a more comprehensive mitigation approach by accounting for control and data dependencies between tasks. 

Hence, adaptation chains provide a balanced response to various considerations, offering flexibility based on violation severity and user preferences for a customized and adaptive approach. Additionally, they excel in managing multi-pronged attacks, efficiently addressing the complexity of sophisticated scenarios. This strategy enhances system resilience and robustness by integrating multiple adaptations, thereby increasing overall adaptability and flexibility.

In conclusion, this paper has established an approach for adapting workflows in response to security violations using RL. Our method, implemented as an extension of jBPM and Cloudsim Plus, demonstrated improved total cost outcomes in response to security violations. 

In future work, we aim to extend our research to address other potential adversaries, such as tenants and their users, and provide security measures against these attackers. This expansion will further bolster the robustness and effectiveness of our proposed approach in ensuring secure cloud-based workflow execution.